# Goos-Hänchen effect in light transmission through bi-periodic photonic-magnonic crystals

Yu. S. Dadoenkova,[1,2,3] N. N. Dadoenkova,[2,3] J. W. Kłos,[4] M. Krawczyk,[4] and I. L. Lyubchanskii[3]

[1]*Institute of Electronics and Information Systems, Novgorod State University, 173003 Veliky Novgorod, Russian Federation*

[2]*Ulyanovsk State University, 432017 Ulyanovsk, Russian Federation*

[3]*Donetsk Physical and Technical Institute of the National Academy of Sciences of Ukraine*

[4]*Faculty of Physics, Adam Mickiewicz University in Poznań, 61-614 Poznań, Poland*

We present a theoretical investigation of Goos-Hänchen effect, i.e. the lateral shift of the light beam transmitted through one-dimensional bi-periodic multilayered photonic systems consisting of equidistant magnetic layers separated by finite size dielectric photonic crystals. We show that the increase of the number of periods in the photonic-magnonic structure leads to increase of the Goos-Hänchen shift in the vicinity of the frequencies of defect modes located inside the photonic band gaps. Presence of the linear magneto-electric coupling in the magnetic layers can result in a vanishing of the positive maxima of the cross-polarized contribution to the Goos-Hänchen shift.

## I. INTRODUCTION

The Goos-Hänchen effect consists in the lateral shift of a reflected light beam with respect to the prediction of geometric optics [1]. Nowadays, this effect is intensively studied [2–7] in different systems, including anisotropic crystals and magnetic materials [8–16], graphene [4, 17–19], superconductors [20, 21], and photonic crystals (PCs) [19–22] despite the long history of investigations since the first observation of this phenomenon in 1947 [1]. Some aspects of Goos-Hänchen effect were studied in different structures (see for example recent review papers [23, 24]). The Goos-Hänchen shift (GHS) of partially coherent light in epsilon-near-zero metamaterials was theoretically investigated in [25]. Giant GHS (up to 70 wavelenghts) and angular shift of several hundred microradians were observed in 2D metasurfaces composed of PMMA gratings on a gold surface [26]. Giant GHS has been predicted also for spin waves in magnetic films [27–30], and optical phonons [31]. This phenomenon has potential application in designing of integrated optics devices, such as optical switchers, bio- and chemical sensors and detectors [32–36].

The GHS in multilayered systems can exhibit interesting peculiarities. For instance, giant GHSs are experimentally demonstrated from a prism-coupled PC structure through a bandgap-enhanced total internal reflection [37]. All-optical tunability of the GHS in PCs was demonstrated in [38]. The lateral shift of the reflected beam can be remarkably enhanced when the phase matching conditions are satisfied for the surface polaritons excitation at the interface of the structure in the graphene-induced photonic band gap (PBG) [39]. The GHS reversibility near the band-crossing structure of PCs containing left-handed metamaterials was shown in [40]. The GHS at the PBG edges can reach values up to several hundreds of wavelengths [41]. Similarly, when a light beam is incident on a PC containing a defect layer, the GHSs are greatly enhanced near the defect mode due to the electromagnetic waves localization [42]. On the contrary, in PT-symmetric crystals the GHS can be huge inside the reflection band [43, 44]. The following bi-periodic structures are also of potential interest: photonic-magnonic crystals (PMCs) which provide PBGs in spectra of electromagnetic waves and magnonic band gaps in spin waves spectra [45–47], and photonic hypercrystals [48].

In this paper, we investigate GHS of light in one-dimensional bi-periodic PMCs. The photonic spectra of such PMCs are characterized by narrow inside-bandgap modes with fine structure [45–47], and it would be expected that the GHS could reach large values in the vicinity of these modes. We take into account magneto-electric properties of the magnetic layers because they can provide important influence on the electromagnetic wave propagation [49].

## II. MODEL AND METHOD OF GOOS-HÄNCHEN SHIFT CALCULATION

We consider PMCs consisting of magnetic layers $M$ of thickness $d_M$ separated by non-magnetic dielectric spacers composed of alternating layers $A$ and $B$ of thicknesses $d_A$ and $d_B$, respectively, as illustrated in Fig. 1. The magnetic layers are magnetically saturated with the magnetization vector **M** lying in the incidence plane and parallel to the interfaces (longitudinal magneto-optical configuration). As a reference structure, we investigate $ABA$ three-layer placed between two magnetic films $M$ (the structure $M(ABA)M$, Fig. 1(a)). Then we increase the number $N$ of dielectric unit cells ($AB$) in the non-magnetic spacer and consider it as a PC of structure $(AB)^N A$ of thickness $d_d = N(d_A + d_B) + d_A$. The magnetic super-cell $[M(AB)^N A]$ is repeated $K$ times (Fig. 1(b)), so that the structure $[M(AB)^N A]^K M$ is bi-periodic PMC [45–47]. We assume a light beam of the angular frequency $\omega$ is incident under angle $\theta$ from vacuum (with (xz) being the incidence plane) and at transmission trough the structure undergoes a lateral shift ΔL. A specific class of magnetic materials possesses spontaneous magneto-electric properties [49]. The magneto-electric effect consisting in a magnetization induction by an electric field and a dielectric polarization induction by a magnetic field was observed in many systems, including magnetic garnets [50, 51]. This effect increases the cross-polarized contribution to the light reflected from a magneto-electric film [52–54] and thus can enhance the corresponding GHS up to several times [15].

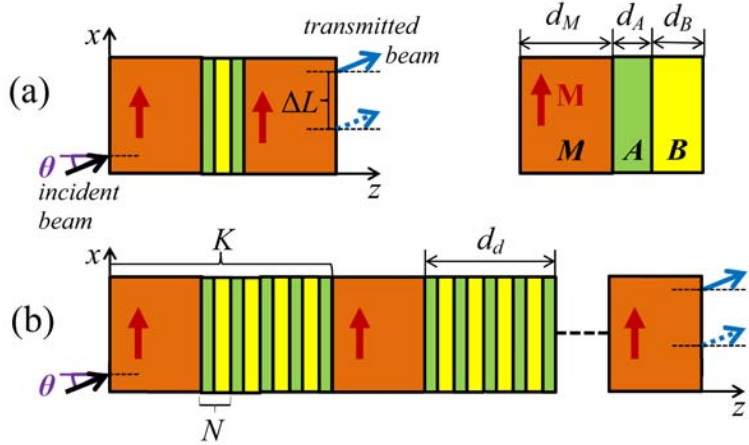

FIG. 1. Schematic of the photonic-magnonic structures consisting of magnetic layers $M$ (thickness $d_M$) and dielectric layers $A$ and $B$ (thicknesses $d_A$ and $d_B$): (a) $M(ABA)M$; (b) $[M\,(AB)^N A]^K\,M$. Thickness of the non-magnetic spacers between the magnetic layers is (a) $2d_A + d_B$, and (b) $d_d = N(d_A + d_B) + d_A$. Red arrows show the magnetization $\mathbf{M}$ direction in the magnetic layers, $\theta$ is the incidence angle of light, and $\Delta L$ is the GHS.

Taking into account the linear magneto-electric interaction, the electric displacement vector $\mathbf{D}^{(M)}$ and the magnetic induction $\mathbf{B}^{(M)}$ in the magnetic layers $M$ are connected with the electric field $\mathbf{E}^{(M)}$ and the magnetic field $\mathbf{H}^{(M)}$ of the electromagnetic wave *via* the constitutive relations [49]:

$$D_k^{(M)} = \varepsilon_0 \varepsilon_{kl}^{(M)} E_l^{(M)} + \alpha_{kl}^{(M)} H_l^{(M)}, \tag{1a}$$

$$B_k^{(M)} = \mu_0 \mu_{kl}^{(M)} H_l^{(M)} + \alpha_{kl}^{(M)} E_l^{(M)}, \tag{1b}$$

where $(k,l) = (x,y,z)$, $\varepsilon_0$ and $\mu_0$ are the vacuum permittivity and permeability, $\hat{\varepsilon}^{(M)}$ and $\hat{\mu}^{(M)}$ are permittivity and permeability tensors of magnetic medium, and $\hat{\alpha}^{(M)}$ is linear magneto-electric tensor which is diagonal in crystals with a cubic symmetry ($\alpha_{kl}^{(M)} = \alpha\,\delta_{kl}$, with $\delta_{kl}$ being the Kronecker symbol) [55]. For bigyrotropic magnetic layers, the non-zero components of $\hat{\varepsilon}^{(M)}$ and $\hat{\mu}^{(M)}$ tensors in the linear magneto-optical approximation are [56]:

$$\varepsilon_{xx}^{(M)} = \varepsilon_{yy}^{(M)} = \varepsilon_{zz}^{(M)} = \varepsilon^{(M)}, \ \varepsilon_{yz}^{(M)} = -\varepsilon_{zy}^{(M)} = i\,\varepsilon', \tag{2a}$$

$$\mu_{xx}^{(M)} = \mu_{yy}^{(M)} = \mu_{zz}^{(M)} = \mu^{(M)}, \ \mu_{yz}^{(M)} = -\mu_{zy}^{(M)} = i\,\mu'. \tag{2b}$$

In the isotropic non-magnetic layers $A$ and $B$, the constitutive relations simply write

$$D_k^{(A,B)} = \varepsilon_0 \varepsilon^{(A,B)} \delta_{kl} E_l^{(A,B)}, \tag{3a}$$

$$B_k^{(A,B)} = \mu_0 \delta_{kl} H_l^{(A,B)}. \tag{3b}$$

We use the (4×4) transfer matrix method [57, 58] to calculate the transmission matrix $\hat{T}$ which components connect amplitudes of the transmitted wave $E_{p,s}^{(t)}$ to those of the incident one $E_{p,s}^{(i)}$ as $E_i^{(t)} = T_{ij} E_j^{(i)}$. Hereinafter the subscripts $(i,j) = (s,p)$ refer to $s$- and $p$-polarizations. The off-diagonal components of $\hat{T}$ correspond to the cross-polarized contribution to the transmission due to bigyrotropic properties of the magnetic layers. Assuming that the incident beam is a Gaussian wavepacket with waist $w_0$ and using the stationary phase method [59], one can derive the GHS $\Delta X_{ij} = \Delta L_{ij} / \lambda$ (in the units of the wavelength $\lambda$) of each component of the transmitted beam in terms of the complex transmission coefficient $T_{ij}$ and its phase $\arg(T_{ij})$, and the $x$-component of the wavevector $k_x$ as:

$$\Delta X_{ij} = -\frac{\partial \arg(T_{ij})}{\lambda \partial k_x} + \frac{\partial \ln|T_{ij}|}{\lambda \partial k_x} \frac{\partial^2 \arg(T_{ij})}{\partial k_x^2} \left( w_0^2 + \frac{\partial^2 \ln|T_{ij}|}{\partial k_x^2} \right)^{-1}. \tag{4}$$

The different components of the transmission matrix $T_{ij}$, and thus the values of the GHS $\Delta X_{ij}$ obtained for any combination of incident and transmitted states of polarization can be separately evaluated and measured in the experiment. Practically, this simply requires the use of a polarizer and an analyzer placed on the path of the incoming and transmitted beams, respectively.

It should be noted that usually only the first term in Eq. (4) is taken into account for calculation of the GHS. However, this approximation is strictly valid for a slowly varying complex transmission coefficient (and resulting phase), wherein, a Taylor series expansion of the transmission coefficient is done and only the first order term (in the k-vector spread) is retained. There has been several reports on more accurate calculations keeping higher order terms (see for example, Ref. [60]) and specifically it has been shown that the magnitude of the shifts deviate significantly where there is an abrupt gradient (such as for angle of incidence close to the Brewster angle). The similar abrupt changes in the transmission coefficients take place in the vicinity of the PBG edges and inside-bandgap modes of the PMCs. Thus, for an improved accuracy of the method, we developed the stationary phase approach by taking into account the second-order term in the Taylor expansion of the phase of the complex transmission coefficient of the PMC.

However, for spatially wide beam ($w_0 >> \lambda$) the second term in Eq. (4) can be neglected even for relatively rapid changes of the transmission coefficients in k-vector domain. It is worth to notice that the condition $w_0 >> \lambda$ must be fulfilled both in experimental and numerical studies when the weak divergence of the beam is required to determine its direction and shift.

## III.   NUMERICAL CALCULATIONS OF THE GOOS-HÄNCHEN SHIFT

For the numerical calculations, we consider the magnetic layers to be of yttrium-iron garnet (YIG) $Y_3Fe_5O_{12}$, which is transparent in the near-infrared regime and possesses bigyrotropic properties. As layers $A$ and $B$ we take titanium oxide $TiO_2$ and silicon oxide $SiO_2$, respectively, with $\varepsilon^{(A)} \approx 5.99$ and $\varepsilon^{(B)} \approx 2.32$ in the considered frequency range [61]. The permittivity and permeability tensor elements of YIG are $\varepsilon^{(M)} \approx 4.81$, $\mu^{(M)} = 1$ [62], $\varepsilon' = -2.47 \times 10^{-4}$, and $\mu' = 8.76 \times 10^{-5}$ [63]. The linear magneto-electric constant of thin epitaxial YIG film is $\alpha = 30$ ps×m$^{-1}$ [50]. The thicknesses of the layers are taken to be $d_M = 700$ nm, $d_A = 190$ nm, and $d_B = 285$ nm, which provides a near-infrared PBG in the transmittivity spectra of the PCs under consideration [45–47]. While fabricating a multilayer structure, the thicknesses of the layers can be distorted. However, a slight change in the thicknesses of the films will not change the behavior of the transmittivity and lateral shifts significantly. This would only lead to a slight drift of the frequency positions of the inside-bandgap modes and PBG edges.

We consider an incident Gaussian beam with waist $w_0 = 30$ μm. In this case the impact of the second term in Eq. (4) to the GHS experienced by the transmitted beam is about 5% in the vicinity of the PBG edges and inside-bandgap modes.

### A. Transmittivity and Goos-Hänchen shift spectra

The transmission spectra $|T_{pp}|$, $|T_{ss}|$, and $|T_{ps}| = |T_{sp}|$ and the corresponding GHSs $\Delta X_{pp}$, $\Delta X_{ss}$, and $\Delta X_{ps} = \Delta X_{sp}$ are presented in left and right panels of Figs. 2, 3, and 4, respectively, for different photonic structures (see Fig.1). Here we neglect the magneto-electric coupling in YIG layers, so that $\alpha = 0$.

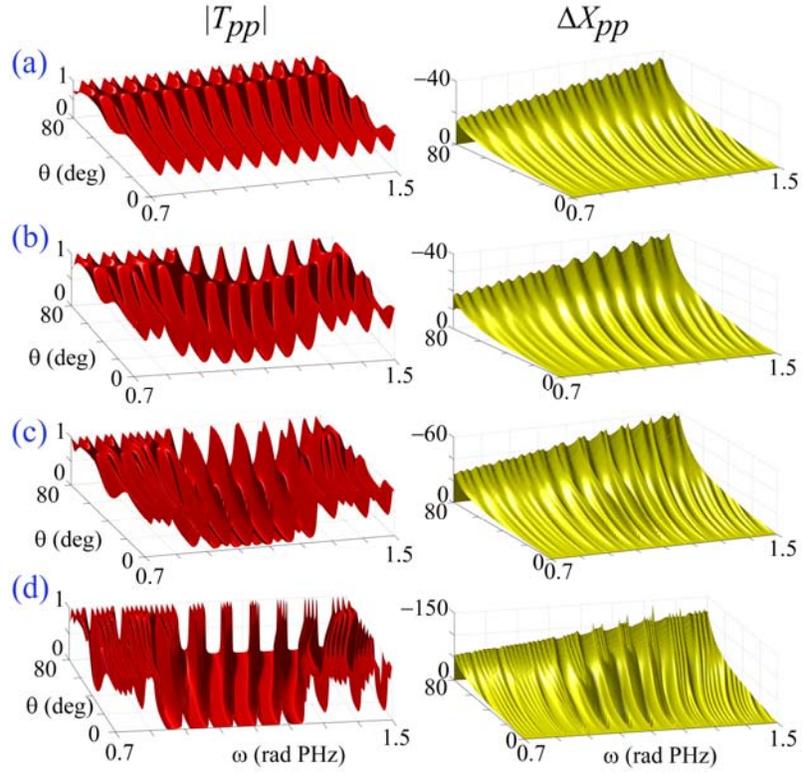

FIG. 2. Transmission coefficient $|T_{pp}|$ and the corresponding dimensionless GHS $\Delta X_{pp}$ as functions of the angular frequency $\omega$ and incidence angle $\theta$ for the structures: (a) $M(ABA)M$; (b) $M\,(AB)^4A\,M$; (c) $[M\,(AB)^4A]^2\,M$; (d) $[M\,(AB)^4A]^5\,M$.

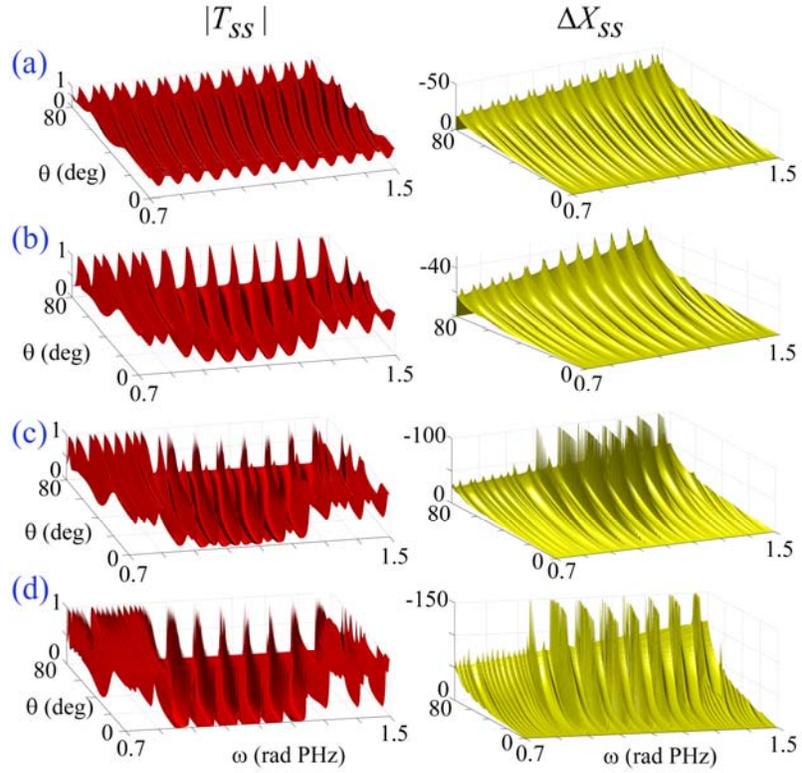

FIG. 3. The same as in Fig. 2 except for $|T_{ss}|$ and $\Delta X_{ss}$.

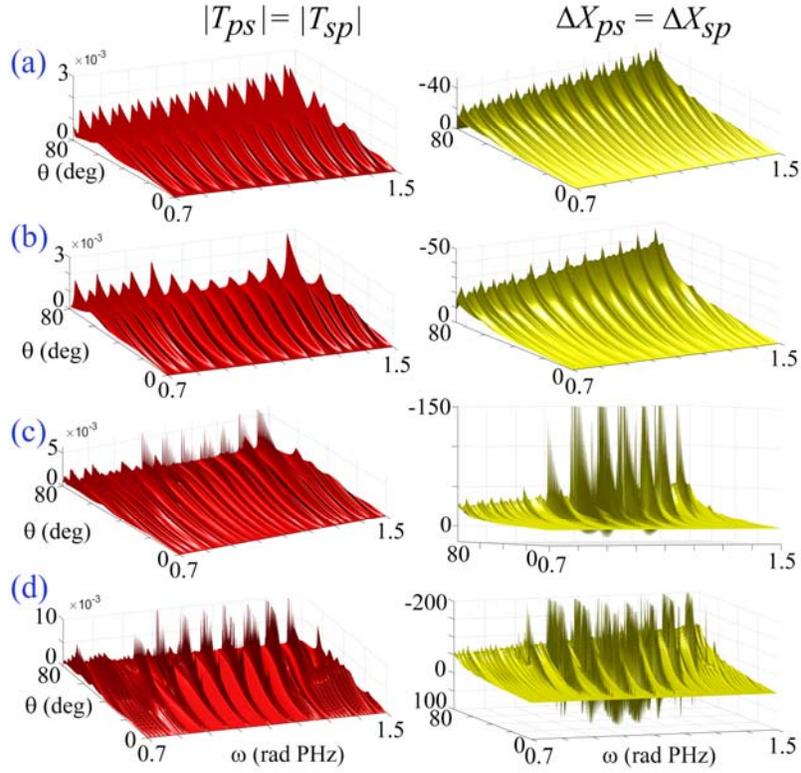

FIG. 4. The same as in Fig. 2 except for $|T_{ps}| = |T_{sp}|$ and $\Delta X_{ps} = \Delta X_{ps}$.

The transmission spectra of the system $M(ABA)M$ (see Fig.1(a)) demonstrate stripe structure of minima and maxima. The values of the diagonal transmission matrix components $|T_{pp}|$ and $|T_{ss}|$ (Figs. 2(a) and 3(a)) vary from 0.1 to 1, while the maxima of the off-diagonal component $|T_{ps}|$ do not exceed $3 \times 10^{-3}$ (Fig. 4(a)). The corresponding GHSs are negative. The diagonal shift $\Delta X_{pp}$ is about several wavelengths for the incidence angles $\theta < 60°$, and its values slowly vary according to minima and maxima of $T_{pp}$, though this variation is about a few $\lambda$ (see Fig. 2(a)). At $\theta > 60°$, $\Delta X_{pp}$ increases in absolute value up to four tens of $\lambda$. For the fixed $\theta$, the values of $\Delta X_{pp}$ increase with the increase of the frequency $\omega$. The GHSs $\Delta X_{ss}$ and $\Delta X_{ps}$ demonstrate the similar tendencies to that for $\Delta X_{pp}$ at $\theta < 60°$, but at $\theta > 60°$ have more pronounced difference in minima and maxima because the variation of $T_{ss}$ and $T_{ps}$ (as well as their phases) responsible for the GHS (Eq. (1)) is higher.

Further addition of $(AB)$ bilayers between the magnetic layers $M$ leads to appearance of the PBG in the transmission spectra. The left (right) panels in Figs. 2(b), 3(b), and 4(b) show $|T_{pp}|$, $|T_{ss}|$, and $|T_{ps}|$ ($\Delta X_{pp}$, $\Delta X_{ss}$, and $\Delta X_{ps}$) for the system $[M(AB)^N A]^K M$ with $N = 4$ dielectric cells and $K = 1$ magnetic super-cell, so that this structure presents a dielectric PC with two defect layers $M$. A set of wide and equidistantly distributed defect modes appears in the PBGs spectra. These modes shift towards higher $\omega$ with the increase of $\theta$ for all polarization states. In the case of $T_{pp}$, the defect modes merge at $\theta \approx 60°$, and the PBG shrinks (Fig. 2(b)). On the contrary, the widths of the PBGs in spectra of $T_{ss}$ and $T_{ps}$ increase with $\theta$, as shown in Figs. 3(b) and 4(b), respectively. Variation of the GHSs at the frequencies inside the PBGs becomes more pronounced relatively to the structure $M(ABA)M$ with a single dielectric period, as follows from comparison of Figs. 2(a), 3(a), 4(a) and 2(b), 3(b), 4(b). Frequency positions of the negative maxima of $\Delta X_{ij}$ correspond to the defect modes positions.

Figures 2(c), 3(c), and 4(c) show results for the structure $[M(AB)^4A]^K M$ with $K = 2$ magnetic super-cells. This structure can be treated as a dielectric PC with three magnetic defect layers $M$. The defect modes in the PBGs spectra overall become narrower and at high $\theta$ do not merge anymore in $T_{pp}$ spectrum. The values of the defect modes maxima of $T_{ps}$ are larger than for the previously described structures, as one can see from comparison of Figs. 4(a), 4(b), and 4(c). The negative maxima of the corresponding GHSs increase and can reach values of about $-100\lambda$ even at low $\theta$.

With the extension of the structure by increasing the number $K$ of super-cell $[M(AB)^4A]$, the magneto-photonic structure can be considered as a PMC with bi-periodicity: a magnetic and a dielectric ones with the periods $d_M + d_d$ and $d_A + d_B$, respectively. The PBG is still formed due to the dielectric PCs $(AB)^4$, and repetition of the magnetic super-cells lead to forming the inside-bandgap modes of high transmittivity. This is illustrated in Figs. 2(d), 3(d), and 4(d) for the PMC with $K = 5$ magnetic super-cells. The positions of the inside-bandgap modes of $T_{pp}$, $T_{ss}$ and $T_{ps}$ coincide only at normal incidence. The intensity of the inside-bandgap modes in $T_{ps}$ spectra is higher than in the previously discussed structures. Moreover, outside the PBG the values of $|T_{ps}|$ increase with $K$ up to $10^{-2}$ (Figs. 4(a) and 4(d)). The negative maxima of the GHSs which correspond to the inside-bandgap modes can reach values up to $-200\lambda$ (Fig. 4(d)). But the peaks of GHS for $s$-$s$ and $p$-$s$ ($s$-$p$) transmission ($\Delta X_{ss}$ and $\Delta X_{ps}$) are sharper and reach ~~higher~~ values higher than for $p$-$p$ transmission ($\Delta X_{pp}$). It should be noted that $\Delta X_{ps}$ in PMCs with $K \geq 2$ possesses positive values at some inside-bandgap modes frequencies (Figs. 4(c) and 4(d)). The increase of the GHSs around the inside-bandgap modes positions is caused by abrupt change of the transmission coefficient's both absolute value and phase. This can be seen from Fig. 5 which shows color maps of the transmittivity $|T_{pp}|$, its phase arg($T_{pp}$) and GHS $\Delta X_{pp}$ evolution with the angular frequency $\omega$ and incidence angle $\theta$ for the case of $p$-$p$ transmission. As was mentioned above, the impact of the second term of Eq. (4) to the calculated GHS does not exceed 5%. Thus the GHS values are mostly provided by the phase variation. Indeed, as Figs. 5(a) and 5(b) show, in the vicinity of the inside-bandgap modes the phase of the transmission coefficient exhibits abrupt change that results in large GHS, illustrated in Fig. 5(c). This behavior is similar for all polarization combinations of the incident and transmitted light.

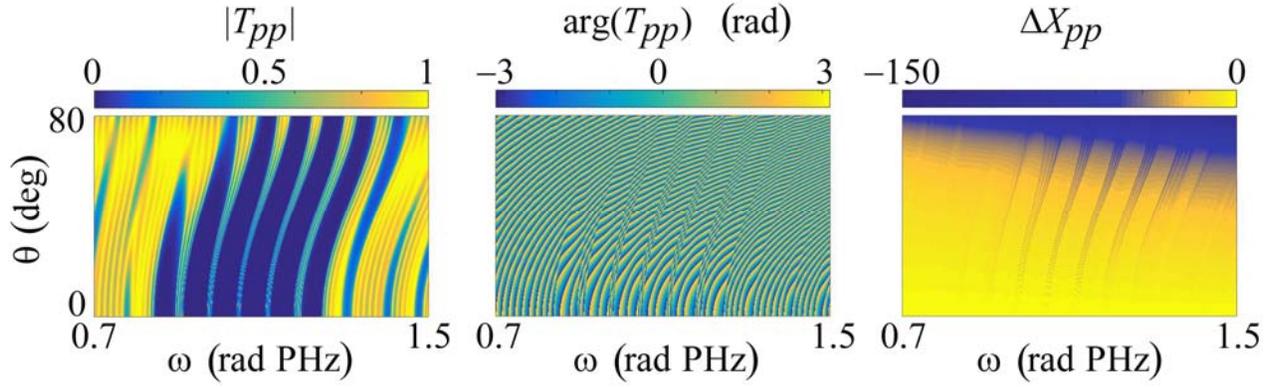

FIG. 5. Color maps of evolution of the transmittivity $|T_{pp}|$, its phase $\arg(T_{pp})$ and GHS $\Delta X_{pp}$ with the angular frequency $\omega$ and incidence angle $\theta$ for the PMC of the structure $[M (AB)^4 A]^5 M$.

### B. Magneto-electric effect influence on the Goos-Hänchen shift in the vicinity of the inside-bandgap mode

As was shown in [45–47], the inside-bandgap modes in the transmittivity spectra of a bi-periodic PMC possess a fine structure, and the number of the sub-peaks is related to the number of the magnetic super-cells. In Fig. 6 we show the influence of the magneto-electric interaction on properties of the light transmitted through the PMC of the structure $[M (AB)^4 A]^5 M$, focusing on the details of a single inside-bandgap mode for $\theta = 30°$. The solid lines correspond to the previously studied case when no magneto-electric interaction is present in the magnetic layers ($\alpha = 0$), and the dotted lines refer to the case when the magneto-electric coupling is taken into account with $\alpha = 30$ ps×m$^{-1}$. The sub-peaks number is the same in each mode in $T_{pp}$ and $T_{ss}$ spectra (blue and green lines in Fig. 6(a)), but it differs from that in $T_{ps}$ spectrum (Fig. 6(b)). The frequency positions of the sub-peaks of $T_{pp}$ and $T_{ss}$ are different but overlap with those of $T_{ps}$ (Figs. 6(a) and 6(b)). The inside-bandgap modes widths in $T_{pp}$ and $T_{ps}$ spectra ($\Delta \omega_{pp, ps} \approx 17$ rad×THz) are larger than that in $T_{ss}$ spectrum ($\Delta \omega_{ss} \approx 10$ rad×THz).

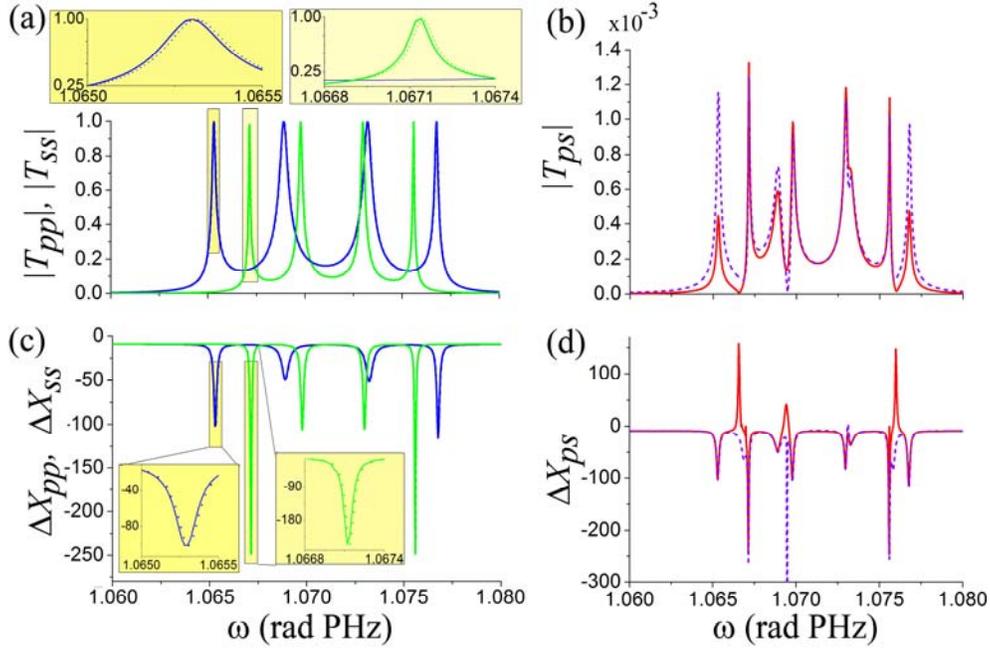

FIG. 6. Fine structure of the inside-bandgap modes in (a) $|T_{pp}|$ (blue lines) and $|T_{ss}|$ (green lines); (b) $|T_{ps}|$; (c) $\Delta X_{pp}$ (blue lines) and $\Delta X_{ss}$ (green lines); (d) $\Delta X_{ps}$ for $\theta = 30°$ and the structure $[M\,(AB)^4A]^5\,M$ in the cases when the magneto-electric constant $\alpha = 0$ (solid lines) and $\alpha = 30$ ps×m$^{-1}$ (dotted lines).

The GHSs demonstrate the same fine structure, and the frequency positions of the sub-peaks of $\Delta X_{ij}$ coincide with those of $T_{ij}$, as shown in Figs. 6(c) and 6(d). As was mentioned above, $\Delta X_{ps}$ can be as negative, as positive, when no magneto-electric interaction is present in the magnetic layers (solid line in Fig. 6(d)), on the contrary to always negative $\Delta X_{pp}$ and $\Delta X_{ss}$. Positions of the positive maxima of $\Delta X_{ps}$ correspond to the position of the gaps between the sub-peaks of $T_{ps}$ (solid lines in Figs. 6(b) and 6(d)). However, not every gap in $T_{ps}$ is accompanied by $\Delta X_{ps} > 0$. For instance, in the middle of the inside-bandgap mode (at $\omega \approx 1.072$ rad×PHz), a smooth change of $T_{ps}$ does not result in large variations of its phase and does not produce a GHS peak.

As one can see from comparison of the solid and dotted lines in the insets in Figs. 6(a) and 6(c), the linear magneto-electric interaction leads to a slight shift of $T_{pp}$ and $T_{ss}$ and the corresponding GHSs $\Delta X_{pp}$ and $\Delta X_{ss}$ towards higher frequencies. However, this shift is only of about 0.02 rad×THz. On the contrary, magneto-electric interaction results in the increase of $T_{ps}$ at low- and high-frequency sub-peaks of the inside-bandgap mode and in a slight modification of the other sub-peaks (Fig. 6(b)). This, in turn, through modification of $T_{ps}$ leads to disappearing the positive maxima of $\Delta X_{ps}$, so the GHS becomes negative in all frequency range (dotted line in Fig. 6(d)).

## IV.   CONCLUSION

In conclusion, we analyzed theoretically the Goos-Hänchen effect of the near-infrared electromagnetic beams in bi-periodic photonic-magnonic crystals and showed a modification of the lateral shift of the transmitted Gaussian wavepacket for all polarization states of the incident and transmitted beams. We showed the increase of the Goos-Hänchen shift at the frequencies of the inside-photonic-band-gap modes and enhancement of the shift peaks due to the linear magneto-electric effect in the magnetic layers for the case of $p$- ($s$-) polarized transmitted beam produced by $s$- ($p$-) polarized incident beam. As shown in recent paper [64], modern experimental set-up allows provide a measurements with high accuracy for different polarization combinations even for relatively weak signal. We hope that the results of our analysis of Goos-Hänchen effect in bi-periodic photonic-magnonic crystals will be useful for the future investigation of complex multiperiodic photonic structures.

## V. ACKNOWLEGEMENTS

This research is supported by: the Ministry of Education and Science of the Russian Federation (Project No. 14.Z50.31.0015 and State Contract No. 3.7614.2017/Π220), the Russian Science Foundation (Project No. 15-19-10036); the European Union's Horizon 2020 research and innovation program under the Marie Skłodowska-Curie (Grant No. 644348), and the Ukrainian State Fund for Fundamental Research (Project No. Φ71/59-2017).

Nanostruct. **11**, 345 (2013).